\documentclass[11pt]{article}
\usepackage{amsmath,geometry}
\usepackage{fullpage}
\usepackage{natbib}
\usepackage{algorithm}
\usepackage{algpseudocode}
\usepackage{lscape}
\usepackage{setspace}
\usepackage{authblk}
\usepackage{graphics,graphicx}
\usepackage{appendix}
\usepackage{subfig}
\usepackage{url}
\usepackage{amsmath,amssymb,amsthm}
\usepackage{bm}
\usepackage{graphicx}
\usepackage{booktabs}
\usepackage{comment}
\usepackage{pdflscape}
\usepackage{rotating}

\newcommand{\betavec}{\boldsymbol{\beta}}
\newcommand{\lambdavec}{\boldsymbol{\lambda}}

\newcommand{\yvec}{\boldsymbol{y}}

\DeclareMathOperator*{\argmin}{arg\,min}

\date{}

\title{The Use of Cross-Validation in the Analysis of Designed Experiments}

\author[1]{\small Maria L. Weese}
\author[2,3,4]{\small Byran J. Smucker}
\author[5]{\small David J. Edwards}
\affil[]{\small Department of Information Systems \& Analytics, Miami University, Oxford, OH}
\affil[2]{\small Henry Ford Health + Michigan State University Health Sciences, Detroit, MI}
\affil[3]{\small Dept of Public Health Sciences, Henry Ford Health, Detroit, MI}
\affil[4]{\small Dept of Epidemiology \& Biostatistics, College of Human Medicine, Michigan State University, East Lansing, MI}
\affil[5]{\small Department of Mathematical Sciences, The Citadel, Charleston, SC}

\begin{document}


\maketitle
\doublespace

\begin{abstract}
Cross-validation (CV) is a common method to tune machine learning methods and can be used for model selection in regression as well. Because of the structured nature of small, traditional experimental designs, the literature has warned against using CV in their analysis. The striking increase in the use of machine learning, and thus CV, in the analysis of experimental designs, has led us to empirically study the effectiveness of CV compared to other methods of selecting models in designed experiments, including the little bootstrap. We consider both response surface settings where prediction is of primary interest, as well as screening where factor selection is most important. Overall, we provide evidence that the use of leave-one-out cross-validation (LOOCV) in the analysis of small, structured is often useful. More general $k$-fold CV may also be competitive but its performance is uneven.

\end{abstract}

\noindent%

\section{Introduction}\label{sec:intro}

\begin{figure}[h]
    \centering
    \includegraphics[width=0.7\linewidth]{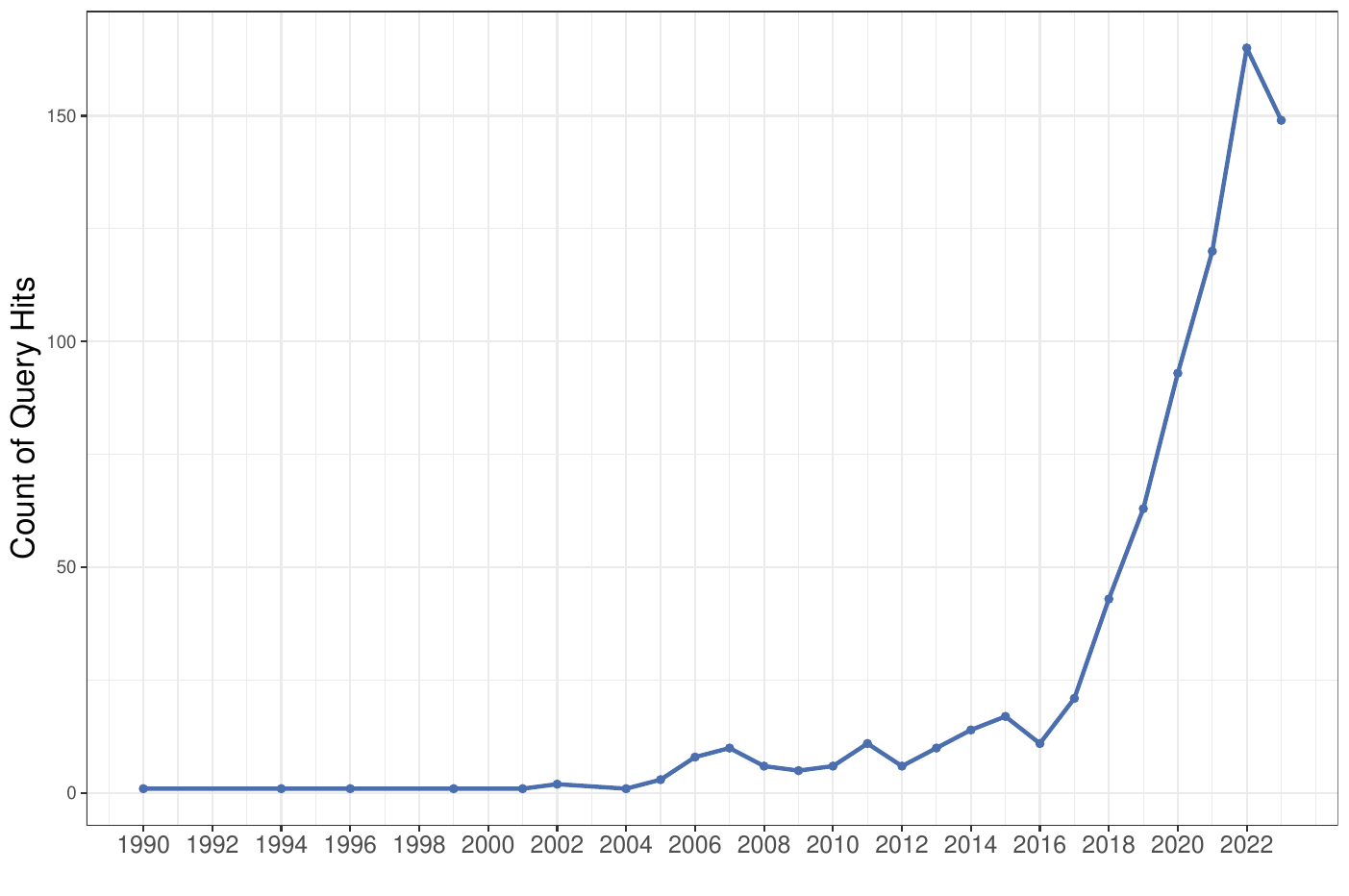}
    \caption{Count of Web of Science citations 1990-2023 for the query: TS=(``Design of Experiment*'' AND ``Machine Learning'')) OR TS=(``Experimental Design'' AND ``Machine Learning'')) NOT TS=(``active learning'')) NOT TS=(``gaussian process'')))) NOT TS=(``sequential experimental design'')) NOT TS=(``hyper-parameter tuning'')) NOT TS=(``hyperparameter*'')) NOT TS=(``Bayesian optimization''))) NOT TS=(``clinical'')) NOT TS=(``clinical trial'')}
    \label{fig:line_plot}
\end{figure}

Two related observations motivate the present work. First, there is a notable increase in the use of machine learning (ML) methods to analyze small designed experiments (DOE+ML). This is evidenced by Figure 6 in \citet{arboretti2022design} 
along with results from our own similar search (Figure \ref{fig:line_plot}).
Secondly, within the design of experiments (DOE) community, it is generally understood that using cross-validation (CV) as an analysis tool in small, structured designs (DOE+CV) will work poorly \citep{yuan2007efficient, draguljic2014screening, lemkus2021self, arboretti2022design}. These two observations are connected because CV is commonly used to tune ML algorithms, even in the cases of such designs. Table \ref{tab:summary} shows a sample of recent papers using DOE+ML, many of which explicitly mention the use of CV. If CV can potentially be used to analyze designed experiments in conjunction with ML, perhaps the received advice not to use CV in the traditional regression-based methods of experiment analysis should be reconsidered as well. 

We say ``potentially'' on purpose in the previous sentence. Although examples proliferate \citep[see][and Table 1]{arboretti2022design}, we are skeptical of much of the ML/DOE synthesis. There is some evidence of success \citep[e.g.,][]{lin2021forward, arboretti2022designQREI}, but we have the sense that the DOE+ML literature is similar to the ``wild west''. For instance, when we tried to reproduce the results in \citet{elkatatny2023optimizing}, we were unable to follow the analytical workflow or achieve results comparable to those reported, despite our efforts and experience. In our view, the expertise required to competently design and analyze a relatively small and highly structured response surface or screening experiment, paired with the complexity, subtlety, and data-neediness of ML methods, will almost certainly lead to many poorly conceived and executed predictive models. The present work is not intended to fully solve this problem, but will provide general insight regarding the wisdom of using cross-validation in the context of experiments.


\begin{table}[H]
	\centering \small
	\begin{tabular}{@{}lcccccccccc@{}}
		\toprule
		Paper & Design 
        & Model 
        & Application \\ \midrule
\cite{escribano2014improvement}* & CCD 
& ANN 
& Welding \\

\cite{raz2018identifying} & D-optimal 
& DNN 
& Information fusion  \\ 

\cite{ghalandari2019aeromechanical}* & Latin Hypercube 
& ANN
& Turbomachinery blades\\

\cite{wiemer2019data}* & Full Factorial 
& DT 
& Engineering applications \\

\cite{pinto2019bootstrap} & CCD/BBD/Doehlert 
& NN 
& Bioprocess development \\


\cite{mathew2020modeling} & BBD  
& ANN
& Supercapacitor optimization \\

\cite{saxena2021battery}* & Half-fraction 
& LASSO/RF 
& Battery Testing \\

\cite{ginige2021solvent} & Full Factorial ($7 \times 4$) 
&  SVM 
& Polymer processing\\

\cite{dropka2022smart} & D-optimal (3-levels) 
& RT 
& Germanium growth \\

\cite{khan2022multi}* & BBD 
&  ANN/SVR 
& Electrochemical dye removal \\

\cite{ratnavel2022predicting} & Taguchi 
& NN 
& 3-D Printing \\

\cite{rebollo2022microfluidic}* & D-optimal Screening 
& ANN 
& Liposomes manufacturing \\
        


\cite{zalkikar2022predictive} & Full Factorial 
& SVM 
& Seawater pipelines \\

\cite{abedpour2023experimental} & Full Factorial 
& ANN 
& Water pollution \\

\cite{elkatatny2023optimizing} & CCD 
& ANN 
& Powder metallurgy \\


\cite{nikita2023process}* & Full Factorial 
& SVR 
& Bioprocess optimization\\

\cite{rabiee2023experimental} & CCD 
& SVR 
& Bone micro-milling \\

\cite{raghavan2023methodology}* & Latin Hypercube 
& RF 
& Laser dicing \\
\bottomrule
	\end{tabular}
 \caption{A summary of a sample of the literature which fits machine learning models using data from experimental designs. Note that in the Model column, we provide the main or preferred ML approach, even if multiple methods were considered; if Model is denoted ``x/y'', neither method clearly preferred. Abbreviations: ANN - artificial neural network; BBD - Box-Behnken design; CCD - central composite design; DNN - deep neural network; DT - decision tree; NN - neural network; RF - random forest; RT - regression tree; SVM - support vector machine; SVR - support vector regression. * indicates the authors specifically mention using CV to train ML models.}
\label{tab:summary}
\end{table}


While prior mention of CV in the context of designed experiments (see citations above) often offer only brief, informal cautions, \citet{breiman1992little} provides perhaps the only theoretically grounded critique of DOE+CV. Breiman demonstrates that when the design matrix is fixed (as is the case in DOE settings), using cross-validation yields more variable prediction error estimates for so-called \textit{unstable} model selection procedures (i.e. those in which a small change in the data can a make large change in the selected model); some examples of unstable procedures include forward selection, all-subsets, tree-based methods, among others. As an alternative to CV in DOE settings, \citet{breiman1992little} proposes a procedure known as the \textit{little bootstrap}, which we describe and examine later in the paper. 

Despite the widespread use of CV in data analysis, no systematic study has examined its behavior in the context of designed experiments. This paper aims to address this gap and evaluate whether Breiman's (and others) concerns are substantiated in practice. More specifically, we will investigate whether cross-validated models offer improved prediction in response surface designs and more accurate model selection in screening scenarios, compared to traditional approaches that avoid CV.  In addition, we compare k-fold CV with leave-one-out-cross-validation (LOOCV) based on the intuition that LOOCV may better preserve the structure of a design.  The little bootstrap will be examined as an alternative to CV altogether.  

The remainder of the paper is organized as follows.  Section 2 presents an overview of the CV algorithm and reviews the literature on its practical implementation and typical use.  We will also review the little bootstrap as a possible alternative to CV.  Section 3 examines the use of CV in response surface modeling and prediction, while section 4 considers its use in the analysis of screening experiments (including supersaturated screening).  Finally, section 5 concludes the article with a discussion and suggestions for future work. 

\section{Cross-Validation}\label{sec:CV}

\begin{figure}
    \centering
    \includegraphics[width=0.6\linewidth]{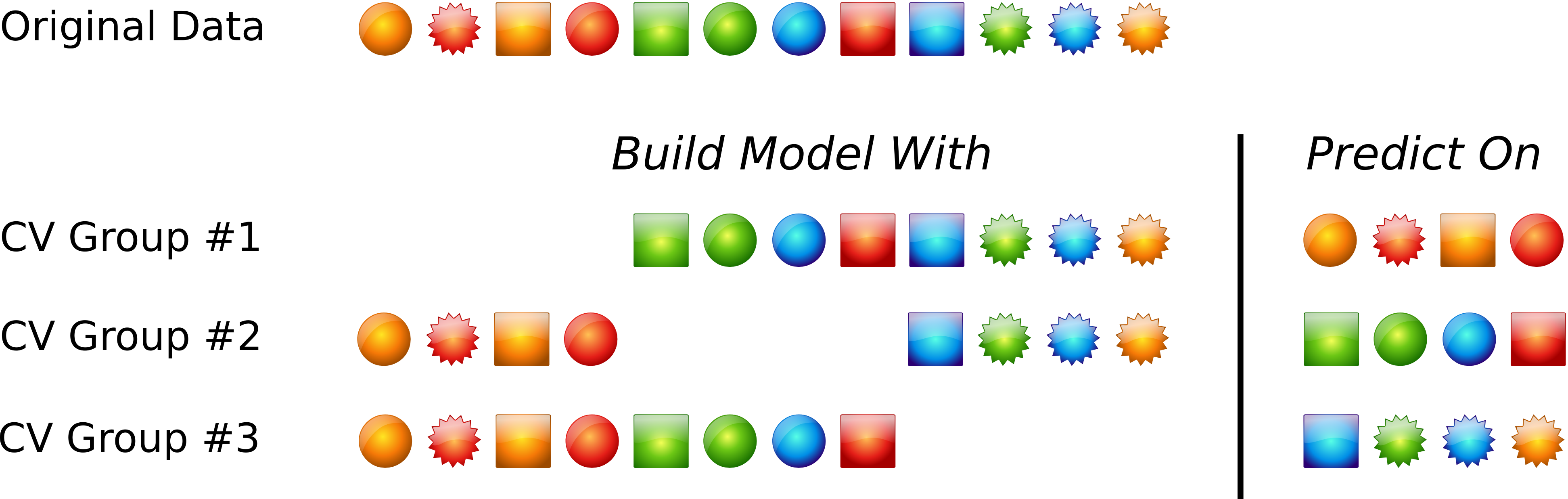}
    \caption{An illustration of $k$-fold CV where $k=3$. Reproduced from \cite{kuhn2013applied}.}
    \label{fig:CV}
\end{figure}

Cross-validation (CV) is an extremely general approach for assessing out-of-sample predictive model performance, based solely on the data at hand. The standard $k$-fold CV procedure, roughly following \citet{james2013introduction}, begins by randomly partitioning the data into $k$ subsets (or folds). The first fold is held out as a validation set, the model is fit on the other $k-1$ folds and then used to predict the validation fold. The quality of this out-of-sample prediction is measured by the root mean squared prediction error, $\text{RMSPE}_{1}=\sqrt{\frac{1}{n^{*}}\sum_{f \in \mathcal{F}_{1}}(y_{f}-\hat{y}_{f})^{2}}$, where $n^*$ is the number of observations in the first fold, $\mathcal{F}_{1}$ is the set of observation indices in the first fold, and $y_{f}$ and $\hat{y}_{f}$ are the $f^{th}$ true and model-predicted value. This process is repeated so that each fold serves as a validation set, producing $\text{RMSPE}_{1}, \text{RMSPE}_{2}, \ldots, \text{RMSPE}_{k}$. The overall $k$-fold CV estimate of out-of-sample prediction error, then, is $\text{RMSPE}_{(k)}=\frac{1}{k} \sum_{i=1}^{k} \text{RMSPE}_{i}$. When $k=n$, the result is leave-one-out cross-validation (LOOCV). Cross-validation provides a reasonable estimate of how the model that produced the predictions $\hat{y}_{f}$ will fare in predicting new observations, and provides an objective method to compare competing predictive models or competing versions of the same predictive model (i.e., models with different hyperparameter settings). Figure~\ref{fig:CV} from \cite{kuhn2013applied} provides a nice toy illustration of 3-fold CV on a dataset with $n=12$ observations and $n^{*}=4$. 

CV has long been a popular approach for assessing the quality of a model \citep{stone1974cross, geisser1975predictive} because it is generally better than using a single holdout and validation set approach \citep{blum1999beating} and more direct than information-based model selection criteria such as AIC \citep{akaikeAIC} and BIC \citep{schwarz1978estimating} that do not explicitly evaluate the model on out-of-sample points. CV has been notoriously difficult to understand theoretically, with extensive discussion in the literature about what quantity CV is actually estimating and how to use its estimate of out-of-sample prediction accuracy to make inferences for predictions \citep[see][and references therein for discussion of both issues]{bates2024cross}. Of more interest to us is its extensive use in model selection and comparison. \citet{arlot2010survey} gives a now-dated summary and \citet{james2013introduction} provides a step-by-step methodology in the context of regression. Many other works have also considered this problem in the context of classical statistical modeling \citep[e.g.,][]{shao1993linear, zhang1993model, ronchetti1997robust, dietterich1998approximate, xu2001monte, yang2007consistency, krstajic2014cross,  zhang2015cross, yates2023cross} and machine learning \citep[e.g.,][]{kohavi1995study, varma2006bias, zhong2010cross, krstajic2014cross, duarte2017empirical, varoquaux2017assessing, a2019evaluation, probst2019tunability, schratz2019hyperparameter, raschka2020modelevaluationmodelselection,     gorriz2024kfoldcrossvalidationbest}. 






Our main goal in this work is to evaluate the use of CV in an experimental design context---that is, to assess DOE+CV. Several previous works considered this in passing only. \cite{yuan2007efficient} apply LOOCV to a small design and conclude it is not as effective as viewing the coefficient path of a penalized regression analysis. Similarly, \citet{draguljic2014screening} found that CV performed poorly for shrinkage models in supersaturated model scenarios. 

\cite{lemkus2021self} also highlights concerns about using CV with designed experiments and proposes a self-validated ensemble modeling approach that uses fractional random weighted bootstrap samples to evaluate predictive performance. 
This is one of a growing number of works exploring the application of DOE+ML. 
Unlike \cite{lemkus2021self}, however, many ML methods use CV for hyperparameter tuning and model evaluation. \citet{arboretti2022designQREI} studied the impact of design choice on the prediction quality of various ML methods, concluding that the MaxPro space-filling design \citep{joseph2020designing} yielded the highest overall quality, with Gaussian process (GP) models providing the best fits for prediction. Interestingly, GPs were the only method tested that did not involve the use of CV. 

Given the apparent lack of CV use in the analysis of experiments, it is fair to wonder if there is a theoretical reason not to use it. \cite{breiman1992little} offers such an argument. He points out that the interpretation of CV as a measure of prediction error assumes that the predictors are randomly sampled from a population. But in the fixed predictor case, as in designed experiments, he argues that CV will overestimate the prediction error. 
In the random predictors case, one assumes that the data, $\{y_n, x_n\}$, are i.i.d. and drawn from some underlying joint distribution $(Y, X)$. We are interested in predicting $y^{new}$ from a model using $x^{new}$ where $(y^{new}, x^{new})$ are also selected independently from $(Y, X)$. But in the fixed predictor case, prediction error is either computed using the same predictor values as in the original design, or in a different portion of the design space that is also known and fixed. 
That is, CV adequately estimates prediction error in the random predictor case, but this estimate is too large in the fixed case. As a consequence, \citet{breiman1992little} proposes an alternative to CV, termed the little bootstrap (LB), for evaluating prediction error in the fixed case. \cite{breiman1996heuristics} further develops the LB as a method to choose the best submodel size from selection procedures like forward selection or best subsets regression.   

Later in this paper, despite Breiman's objections, we evaluate CV in the context of several types of small experiments. But we also include the LB as an alternative. Our implementation is described here, following \cite{breiman1996heuristics}. Let $\mathbf{y}=X\boldsymbol\beta + \boldsymbol\epsilon$ be the standard linear model with $\mathbf{y}$ an $n \times 1$ vector, $X$ an $n\times p$ model matrix, $\boldsymbol\beta$ a $p\times 1$ vector of parameters, and $\boldsymbol\epsilon$ an $n\times 1$ random vector with variance $\sigma^{2}$. The LB procedure is performed as follows:

\begin{enumerate}
\item Fit the full regression model, using $X$, to obtain $\hat{\sigma}^{2}$, an estimate of $\sigma^2$. An alternative estimate of $\sigma^2$ when $p > n$ will be discussed in Section~\ref{sec:Screening}. 
\item For each model size $s$, perform best subsets regression to identify the best-fitting model and compute its residual sum of squares $RSS_s=\boldsymbol y '(I-H_s)\boldsymbol y$, where $I$ is the $n\times n$ identity matrix and $H_s=X_s(X_s'X_s)^{-1}X_s'$ is the projection matrix for the selected model matrix $X_s$. 
\item Generate $\tilde{\boldsymbol\epsilon}$ ($n\times 1$) where each $\tilde\epsilon_i\sim N(0,t\hat{\sigma}^2)$ and $t$ is a constant typically chosen between 0.6 and 1.  We choose $t=0.6$ following \citet{breiman1992little}. 
\item Create a perturbed response value $\tilde{\boldsymbol y}=\boldsymbol y +\tilde{\boldsymbol{\epsilon}}$.
\item For each subset, $s$, refit the selected model using $\tilde{\boldsymbol y}$ and calculate fitted values, $\hat{\tilde{\boldsymbol y}}_s = H_s\tilde{\boldsymbol{y}}$.
\item Calculate the bias term $B=\frac{1}{t^2}\tilde{\boldsymbol \epsilon}' \hat{\tilde{\boldsymbol y}}_s$.
\item Repeat steps 3-6 $n_{\text{bootstrap}}$ times, where $n_{\text{bootstrap}}=25$, typically.  
\item For each submodel size, $s$, average the $n_{bootstrap}$ values of $B$ yielding $\bar{B}_s$
\item Calculate $PE_s=RSS_s+2\bar{B}_s$. The optimal $s$ corresponds to the minimum value of $PE_s$.    
\end{enumerate}

\section{The Use of Cross-Validation in the Analysis of Response Surface Experiments}\label{sec:pred}

Cross-validation is typically used to control model complexity and prevent overfitting, by providing a reasonable assessment of out-of-sample predictive quality. Thus, it is natural to consider its use in response surface experiments, which were designed to be used in the latter stages of an experimental campaign \citep{Box1951, box2007response, myers2016response} when it is clear which numerical factors are important and the experimenter wishes to estimate the values of the factors that optimize a response of interest. That is, an overarching goal of these experiments is good prediction and thus it is worth considering whether cross-validation can help achieve this objective even though the designs are relatively small and structured. \cite{arboretti2022design} provides a thorough comparison of design types and machine learning methods, and found that the Gaussian Process model, a method for which CV was not used, was the best method in terms of out-of-sample prediction. Due to this and doubtful statements in the literature, our expectation was that using CV in the analysis of these experiments would be detrimental to prediction quality. In order to assess this informal hypothesis, we use several analysis methods---concentrating on regression---along with similar designs to those of \cite{arboretti2022design}. We focus mainly on regression analysis because it is the established analysis method for response surface designs and thus can simplify our attempt to isolate the effect of CV.

In this section, we present simulations used to assess the prediction quality of a variety of RSM analysis methods---most of which use CV---over a variety of RSM designs. 

\subsection{Preliminaries} \label{sec:sim_prelim}

The design types are as follows:
\begin{enumerate}
    \item A face-centered central composite design, denoted CCD, and including 7 replicated center points;
    \item A central composite design with axial runs, denoted CCD(axial), and 7 replicated center points, with an axial distance of $\alpha=\sqrt{m}$, where $m$ is the number of factors;
    \item An I-optimal design for the full second-order model.  This design contains 3 replicated center points. 
    \item A Box-Behnken design, denoted BBD, including 3 replicated center points;
    \item A Maximum Projection (MaxPro) space-filling design \citep{joseph2015maximum}.
\end{enumerate}

\noindent From among these design types, we use two different numbers of factors: $m=6$ ($n=52$) as used in \citet{arboretti2022machine} and a smaller design with $m=3$ ($n=16$).  The designs are included in the supplementary material.

For analysis, we focus on regression methods but also add a random forest model, commonly used in the machine learning and DOE literature. More specifically, we compare the following RSM predictive models:
\begin{enumerate}
    \item The full second-order model:
    \begin{align}
        E(Y_{h})=\beta_{0} + \sum_{i=1}^{m} \beta_{i}X_{ih} + \sum_{i=1}^{m}\beta_{ii}X^{2}_{ih} +
        \sum_{i<j} \beta_{ij}X_{ik}X_{jh}. \label{eq:2nd}
    \end{align} 
    That is, we fit model \eqref{eq:2nd} using least-squares and use the resulting fitted regression as a predictive model.
    \item Following Chapter 6 of \citet{james2013introduction}, the least-squares model retaining only the terms of the full second-order model identified via 5-fold CV.
    \item The same procedure as 2., except with LOOCV.
    \item The least-squares model retaining only the terms of the full second-order model using LB as outlined in Section~\ref{sec:CV}.  
    \item A random forest model with 5-fold CV as described in \cite{arboretti2021machine}.
    \item A random forest model with LOOCV.
\end{enumerate}

\noindent Notes on the methods:
\begin{itemize}
    \item Fitting a full second-order regression model 
    is the standard way to analyze an experiment from a response surface design.  
    \item In \citet{james2013introduction}, CV is paired with best subsets regression to develop a predictive regression model, in a way that takes care not to evaluate predictive quality using data that has been used to train the model. First, each run in the design is randomly allocated into one of $k$ folds. For each of the $k$ CV splits, best subsets regression is performed on the $k-1$ training folds. Within each split, for each submodel size $s$, the best-fitting model of size $s$ is selected based on the training data, and its prediction accuracy is evaluated on the corresponding validation fold using the root mean squared prediction error (RMSPE). 
   For example, when $k$=5 folds and the design has $n$ runs, then $n_{train}=\frac{4n}{5}$ will be the approximate size of each of the 5 training samples and $n_{val}=\frac{n}{5}$ will be the approximate size of the validation samples. 
   
   Best subsets regression is run on each of the $k$ training sets of size $n_{train}$. For each of the $s$ best subset models created on each of the $k$ training sets, $\text{RMSPE}_{(k,s)}$ is calculated using the $n_{val}$ validation values ($y_f$) as described in Section~\ref{sec:CV}.  This creates a $k \times s$ matrix with values of $\text{RMSPE}_{(k,s)}$.  Each of the $s$ columns are averaged over the $k$ folds to produce $\overline{\text{RMSPE}}_s$.  The optimal submodel size, $s^*$, is chosen as $s^* = \argmin_s \overline{\text{RMSPE}}_s$. 
    Finally, we select a final model of size $s^*$ by performing best subsets regression on the full data set. This procedure is repeated for LOOCV except we use $k=n$. Best subsets regression is implemented using the \emph{leaps} package in \citep{lumley2013package}. R code for this implementation is provided in the Supplementary Materials and can also be found in Chapter 6 of \cite{james2013introduction}. 
    \item The procedure for using LB to choose the optimal submodel size is outlined in Section~\ref{sec:CV}.  We set $t=0.6$ and use $n_{bootstrap}=25$ \citep{breiman1992little}.  Once the optimal submodel size, $s^*$ is determined via LB, we select a final model of size $s^*$ by performing best subsets regression on the full data set.  
    \item The random forest model was tuned using both 5-fold CV and LOOCV over a simple hyperparameter grid. The search included $mtry$ (the number of columns randomly selected at each split), the splitting rule, and the minimum node size, with tuning aimed at minimizing the average CV or LOOCV prediction error. Further details regarding the hyperparameter grids can be found in the Supplementary Materials. The values of $mtry$ and the minimum node size were adjusted based on $n$ and $m$.  We used the \emph{ranger} package \citep{wright2015ranger} in R which provides a more flexible implementation of the random forest algorithm with fewer hyperparameters. Although \cite{arboretti2022design} found random forests to be suboptimal for prediction, we included them as a baseline machine learning method that uses the out-of-sample prediction error for tuning.
\end{itemize}

Simulation was used to evaluate the $\text{RMSPE}$ for each analysis method and design type. This was done for three different true models, generated using a similar approach to \cite{smucker2021response}. Simulating realistic response surfaces is more complex than simulating general regression data. Thus, response surfaces were generated using the testbed approach of \citet{mcdaniel2000response}, which allows control over properties like effect sparsity, heredity, steepness/flatness. First, we generated a second-order surface with all main effects, two-factor interactions and quadratic terms active. Second, we created a second-order surface where only a fraction of the main effects, two-factor interactions, and quadratic effects are active; active effects are selected based on the \cite{ockuly2017response} probabilities. Finally, we generate a sixth-order response surface including all pure polynomial terms up to the sixth order, along with all two and three-factor interactions. In our simulation, the second-order surfaces were adjusted to be relatively ``flat'', while the more complex sixth-order surface was adjusted to be ``steeper''.  Further details on the testbed and its implementation can be found in the appendices of \cite{smucker2021response}.  The simulation was carried out as follows. For each design type and size:
\begin{enumerate}
    \item Create the specified true response surface using the specified design and an error standard deviation of $\sigma=1$.
    \item Generate 1000 uniformly distributed out-of-sample (new) points on $U(-1,1)$. Obtain true $y_{new}$ values using the same generated response surface as in step 1.
    \item Apply one of the analysis methods described above to identify the ``best'' model.  
    \item Evaluate the selected model on the 1000 out-of-sample points and calculate $\text{RSMPE}=\sqrt{\frac{1}{1000}\sum_(y_{new}-\hat{y}_{new})^2}$.
    \item Repeat each of steps 1-4 500 times for each analysis method. 
\end{enumerate}

\subsection{Simulation Results} \label{sec:pred_sim_results}

\begin{figure}[H]
    \centering
    \includegraphics[width=0.75\linewidth]{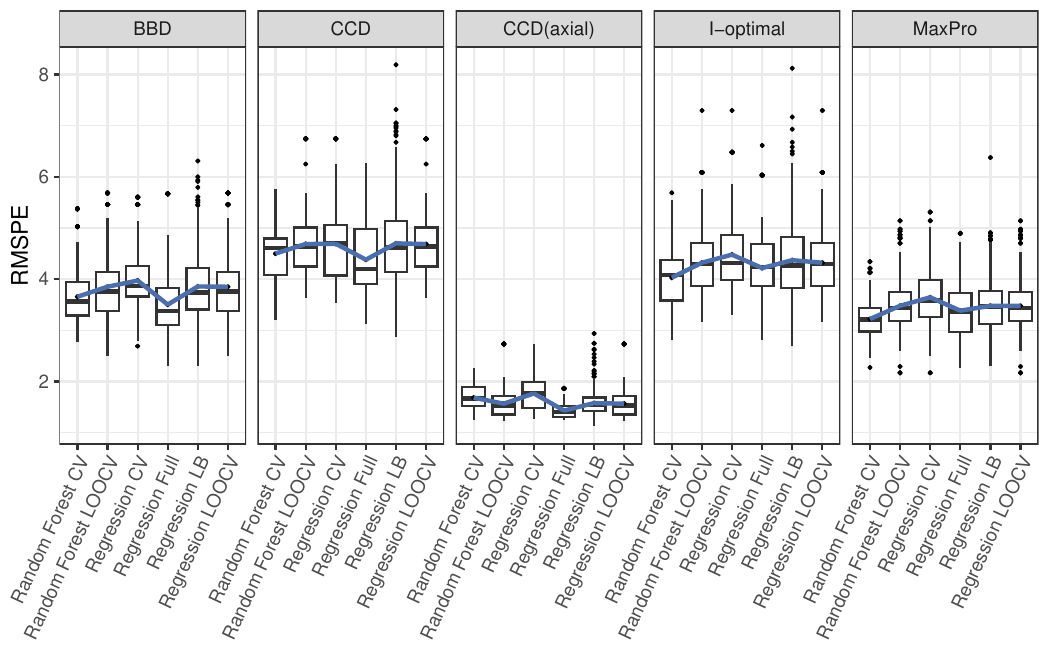}
    \caption{RMSPE for all design types and all analysis methods for a true response surface that is sixth order for $n=52$, $m=6$.  The blue line indicates the average value.}
    \label{fig:preds_sixth}
\end{figure}

Figure~\ref{fig:preds_sixth} displays the RMSPE values for the models fit to the sixth-order data, for $n=52$ and $m=6$. The full second-order regression model, along with the random forest model with CV, exhibit the highest prediction quality, depending on the design type. Regression methods using CV, LOOCV or LB do not seem to differ substantially with respect to RMSPE. The CCD with $\alpha=\sqrt{m}$ displayed the best prediction quality; this is not surprising given the improved precision for estimating curvature when $\alpha > 1$. For designs constrained to $[-1,1]$, the MaxPro design was the best performer with the Box-Behnken design following very closely behind. 

\begin{figure}[H]
    \centering
    \includegraphics[width=0.7\linewidth]{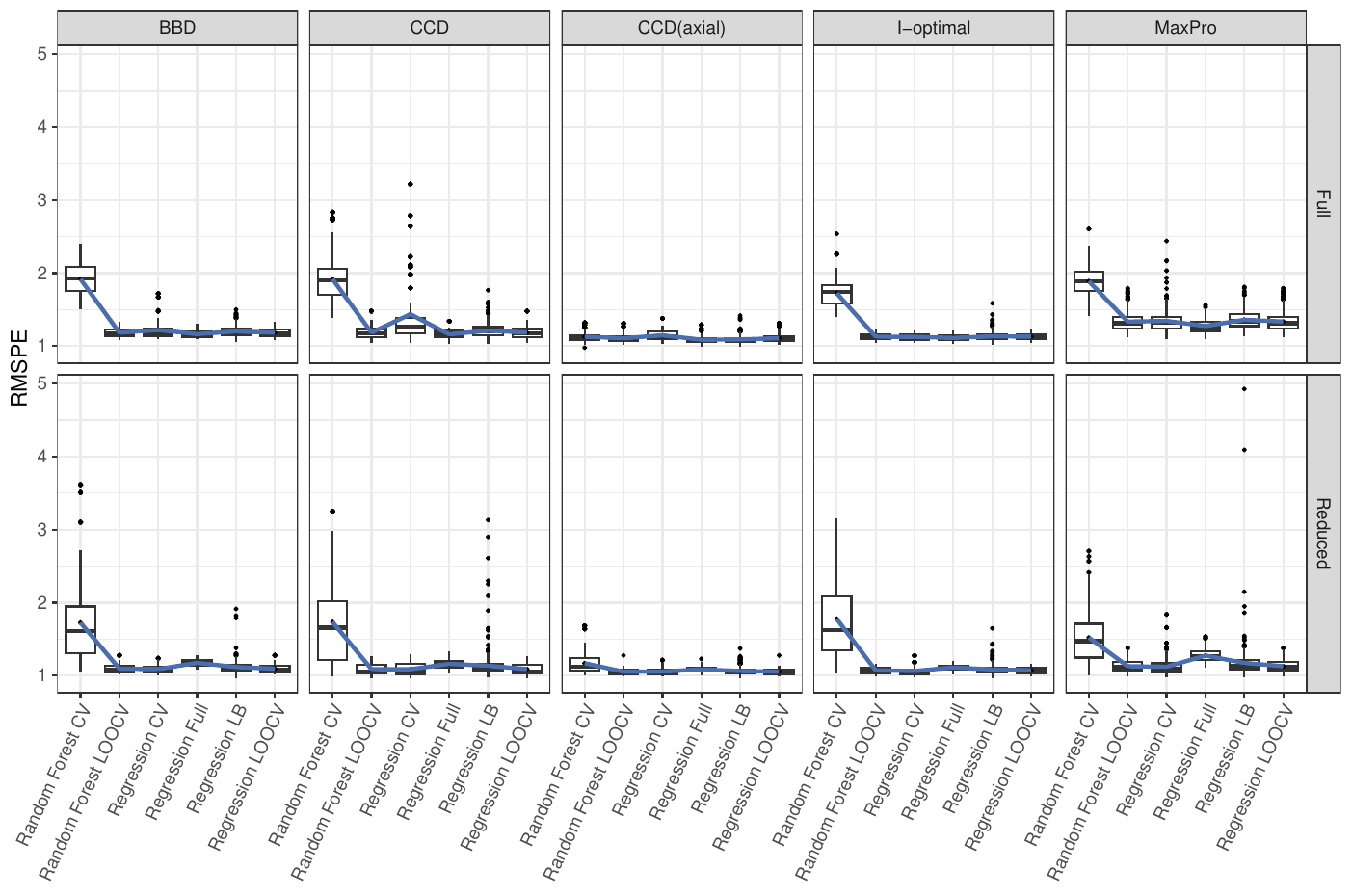}
    \caption{RMSPE for all design types and all analysis methods for a true response surface that is either a full second order surface (Full) or a second order surface that varies the number of true effects, $n=52$, $m=6$. The blue line indicates the average value.}
    \label{fig:preds_second}
\end{figure}

\begin{figure}[H]
    \centering
    \includegraphics[width=0.7\linewidth]{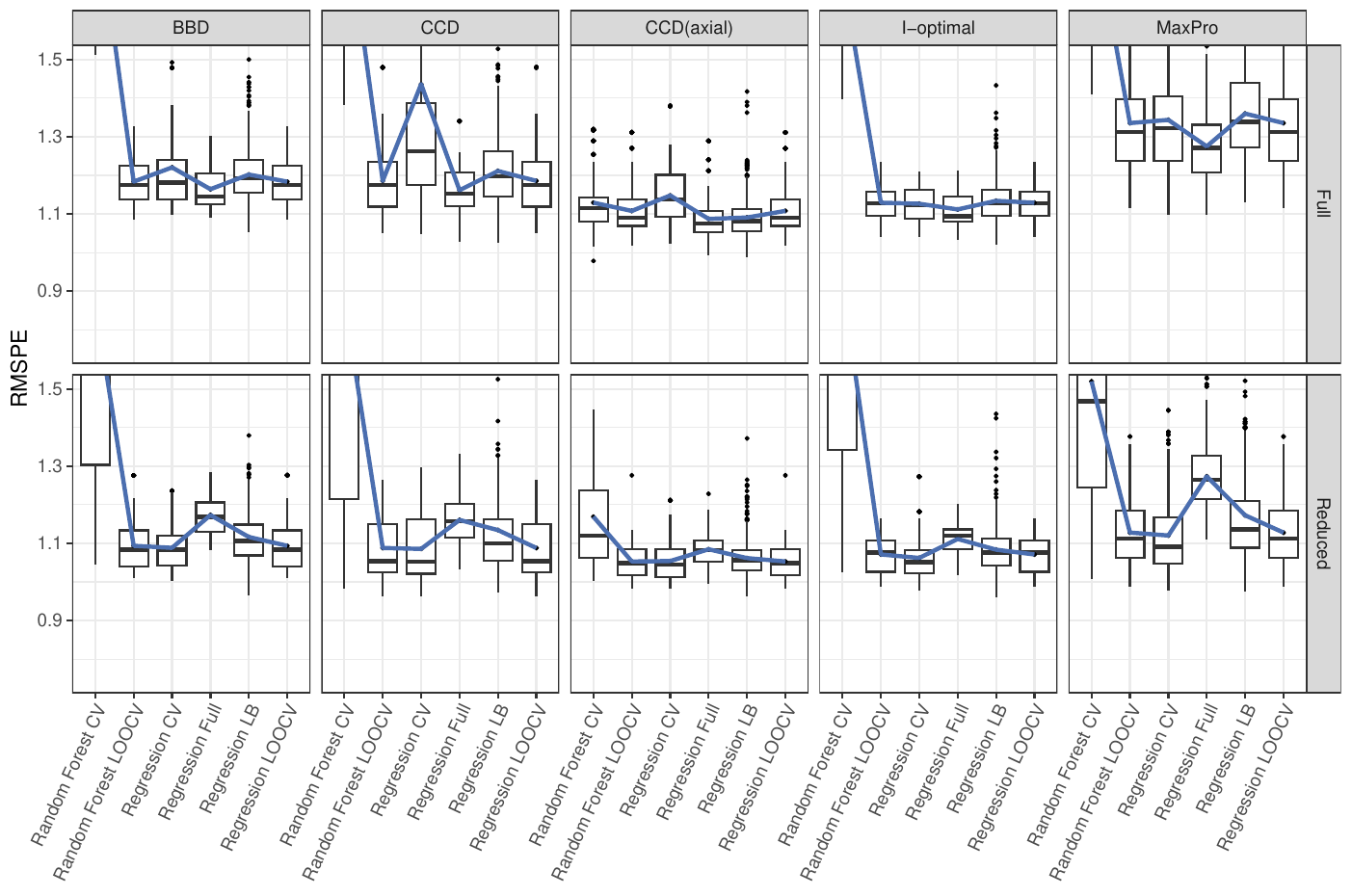}
    \caption{Y-axis zoom of the RMSPE for all design types and all analysis methods for a true response surface that is either a full second order surface (Full) or a second order surface that varies the number of true effects, $n=52$, $m=6$. The blue line indicates the average value.}
    \label{fig:preds_second_zoom_}
\end{figure}

Figure~\ref{fig:preds_second} shows that for true second-order surfaces, all designs (with one exception) indicate the random forest model with 5-fold CV to be overfitting. 
A zoomed version of these results (Figure~\ref{fig:preds_second_zoom_}) shows the random forest model (LOOCV) performs quite competitively with the other methods, across design types. Unsurprisingly, the full regression model performs well when the true model is full second-order and is less effective when the true model is a subset of this model. When the true model is the full second-order model, regression with 5-fold CV performs relatively poorly on CCDs. However, it remains competitive across all designs when the true model is a reduced second-order model. Methods using LOOCV generally perform well, while the LB approach does not appear to provide any clear advantage.


\begin{figure}[H]
    \centering
    \includegraphics[width=0.8\linewidth]{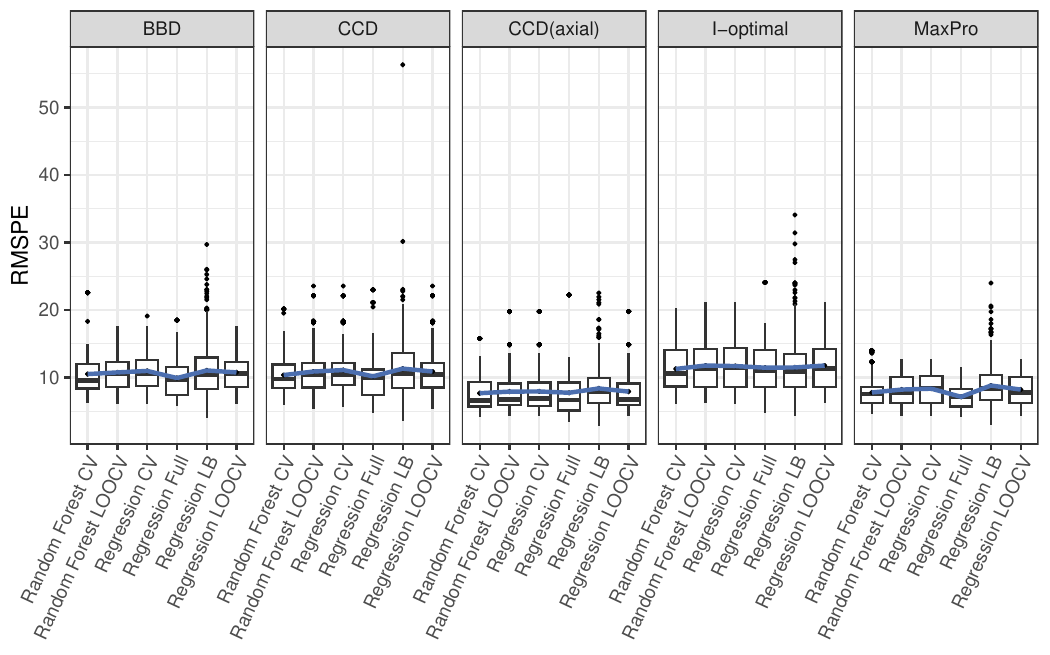}
    \caption{RMSPE for all design types and all analysis methods for a true response surface that is sixth order for $n=16$, $m=3$. The blue line indicates the average value.}
    \label{fig:preds_sixth_16}
\end{figure}

We also evaluated CV for a set of smaller response surface designs, which intuitively should present problems for this method. Figure~\ref{fig:preds_sixth_16} shows RMSPE vales for the $n=16$ and $m=3$ designs when the data is generated from a sixth-order surface. We note few differences between the analysis methods, though RMSPE values are more than double those of the larger designs in Figure \ref{fig:preds_sixth}. There are potentially minor differences, however.  For example, the LB method shows slightly greater variability in performance compared to the others, and fitting the full second-order model appears to yield marginally better prediction accuracy.

\begin{figure}[H]
    \centering
    \includegraphics[width=0.8\linewidth]{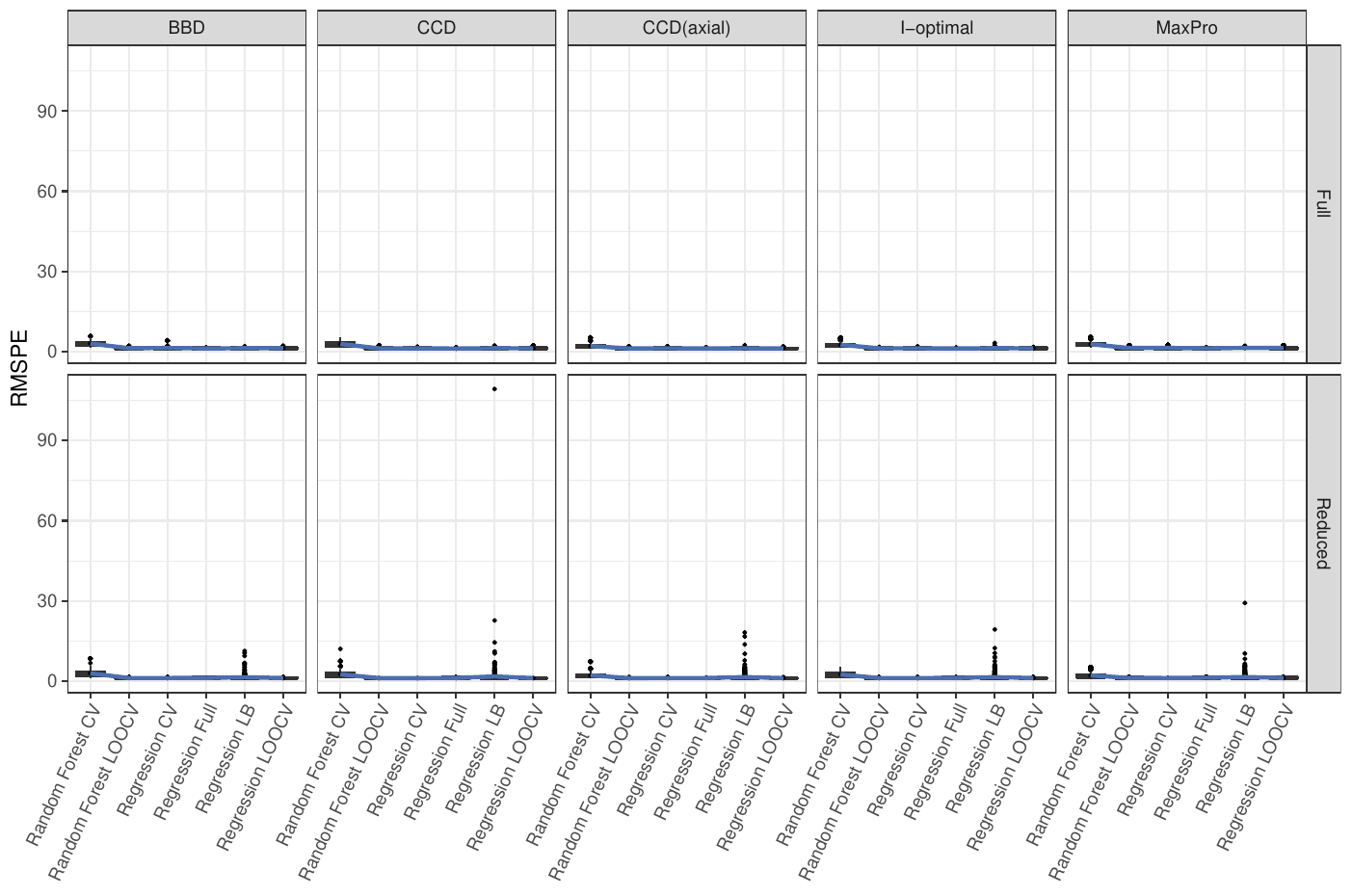}
    \caption{RMSPE for all design types and all analysis methods for a true response surface that is either a full second order surface (Full) or a second order surface that varies the number of true effects, $n=16$, $m=3$. The blue line indicates the average value.}
    \label{fig:preds_second_16}
\end{figure}

\begin{figure}[H]
    \centering
    \includegraphics[width=0.8\linewidth]{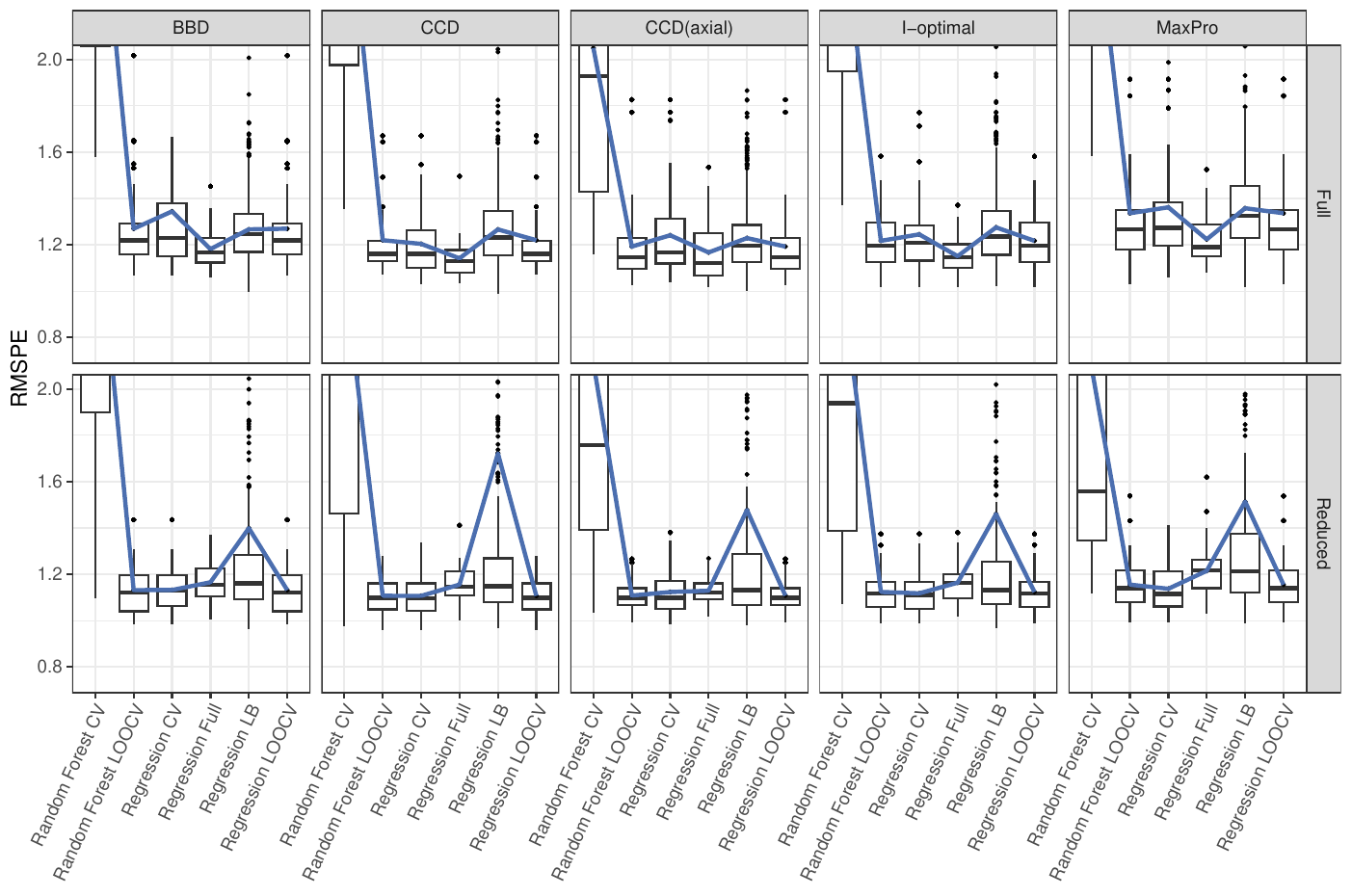}
    \caption{Y-axis zoom of the RMSPE for all design types and all analysis methods for a true response surface that is either a full second order surface (Full) or a second order surface that varies the number of true effects, $n=16$, $k=3$. The blue line indicates the average value.}
    \label{fig:preds_second_zoom_16}
\end{figure}

Figure~\ref{fig:preds_second_16} shows RMSPE values for the full and reduced second-order surfaces; we immediately notice an extreme outlier for the LB/CCD/Reduced combination.  Figure~\ref{fig:preds_second_zoom_16} shows a zoomed-in comparison of the methods.  As with the larger designs, we notice the random forest with 5-fold CV performs poorly, whereas the LOOCV version is quite competitive with the other methods. Across all designs, LB offers no predictive advantage over cross-validated regression.  There is little difference in performance between regression with 5-fold CV or LOOCV (although LOOCV exhibits less variability in RMSPE values for several cases). 

To provide additional insight,  we end the section with some comments on the size of the models selected by the regression methods (Table \ref{tab:model_sizes}). Regression using LB has a larger average model size for the sixth-order surface for both design sizes. For the full and reduced second-order surfaces, the model sizes are more similar. 

\begin{table}[h!]
\centering
\begin{tabular}{lccc}
\toprule
\textbf{Design / Surface} & \textbf{LB} & \textbf{CV} & \textbf{LOOCV} \\
\midrule
$n=16$ / sixth-order & 3.3 & 2.4 & 2.6 \\
$n=52$ / sixth-order & 7.6 & 3.4 & 5.4 \\ 
\hline
$n=16$ / second-order & 6.7 & 6.7 & 7.0 \\
$n=52$ / second-order & 17.1 & 17.5 & 18.7 \\
\hline
$n=16$ / reduced & 3.5 & 3.5 & 3.7 \\
$n=52$ / reduced & 5.0 & 7.8 & 6.5 \\
\bottomrule
\end{tabular}
\caption{Average number of terms chosen by three regression methods, across design sizes and true model surfaces.}
\label{tab:model_sizes}
\end{table}

\section{Cross-Validation in the Analysis of Screening Designs}\label{sec:Screening}

Screening to find the few important factors from among the many unimportant ones is a basic discovery task for which experimentation can be used. In screening experiments, the ratio of runs to factors is much smaller than in response surface studies, and the goal is active effect identification rather than prediction. For these reasons, we might be even more suspicious that CV will be an effective method to aid in model selection. Here we compare the quality of various analysis methods---some involving CV and some not---for the screening task, where the goal is to correctly detect truly active effects without falsely identifying spurious ones. To accomplish this, we will first examine design for which $n>m$, using the designs, simulation scenarios, and simulation protocol of \cite{mee2017selecting}. Then we will study supersaturated designs, for which $n<m$, taking our cue from \citet{marley2010comparison}.  

\subsection{Screening with Nonregular Fractional Factorial Designs}

\cite{mee2017selecting} study multiple measures of design quality, including the ``power'' (the proportion of truly active factors labeled as active) and the type I error rate (proportion of truly inactive factors labeled active). The authors focus on screening designs with $n=20$ runs and $m=7$ factors, assuming an experimental goal of identifying active main effects as well as some two-factor interactions. \citet{mee2017selecting} evaluated two orthogonal arrays (so-called OA18 and OA1a), two types of model-robust designs (the MEPI criterion of \citealp{li2000model} and the PEC criterion of \citealp{loeppky2007nonregular}), and the Bayesian D-optimal designs of \citet{dumouchel1994simple}. 

The simulation protocol of \cite{mee2017selecting} chooses, for each scenario, a fixed number of active main effects and two-factor interactions.  The true models contain 2 or 4 truly active main effects (ME) and 2 or 4 truly active two-factor interactions (2FI). The true regression coefficients ($\boldsymbol\beta$) for the MEs and 2FIs are obtained by sampling with replacement from \{$2,2.5,3,3.5$\}, then randomly assigning $+$ or $-$.  Active two-factor interactions are selected based on the assumption of weak effect heredity; that is, only interactions with at least one active parent main effect are considered.  The true response is generated according to $y=X\betavec+\boldsymbol\epsilon$ where $\boldsymbol \epsilon\sim N(0,I)$. 

We analyzed 250 response vectors for each scenario of active main effects and two-factor interactions using the following analysis methods:

\begin{enumerate}
  
    \item Gauss-Lasso as implemented with a data-driven threshold of $\gamma=0.1*\text{max}|\hat{\betavec}|$ and the BIC statistic to select the value of the tuning parameter $\lambda$ (see below for details).
    \item Lasso using 5-fold CV or LOOCV to choose the value of $\lambda$. 
    \item Regression as described below using either CV or LOOCV.
    \item Regression as described below using LB.
\end{enumerate}

\noindent Notes on the methods:
\begin{itemize}
\item The Gauss-Lasso procedure provides a non-CV alternative to choose the Lasso tuning parameter $\lambda$. It is implemented as follows. Center and scale $X$ and center $\mathbf{y}$, as in \citet{weese2021strategies} and call them $X_{cs}$ and $\mathbf{y}_{c}$, respectively. For each $\lambda \in \lambdavec$, obtain the lasso estimates $\hat{\betavec}=\argmin_{\betavec} \frac{1}{2n}(\yvec_{c}-X_{cs}\betavec)^T(\yvec_{c}-X_{cs}\betavec)+\lambda \left\|\betavec \right\|_1$. For each $\hat\beta_j \in \hat\betavec$ at the $i^{th}$ value of $\lambda$ (denote as $\hat\beta_{\lambda(i,j)}$), a threshold, $\gamma$, is applied such that if $|\hat{\beta}_{\lambda(i,j)}|<\gamma$  then $\hat{\beta}_{\lambda(i,j)}=0$. After thresholding, the columns of $X$ with $\hat\beta_{\lambda(i,j)}>0$ are used to fit a linear model and the corresponding BIC statistic is calculated; BIC$_\lambda = n(\log(RSS_\lambda/n))+p\log(n)$.  The optimal $\lambda^*$ is selected as $\lambda^*=\argmin_\lambda \text{BIC}_\lambda$. The effects associated with values of $\hat{\beta}_{\lambda^*(i,j)}>0$ are declared active.  
R code for the implementation of Gauss-Lasso can be found in the supplementary materials.
\item Lasso with 5-fold cross-validation or LOOCV is implemented using the R package \emph{glmnet} \citep{friedman2021package}, which is used to identify the optimal tuning parameter $\lambda^*$ and select the corresponding nonzero coefficients as active effects.
\item Regression with 5-fold CV or LOOCV is implemented as described in Section~\ref{sec:pred}.  As before, 5-fold CV or LOOCV is used to determine the optimal submodel size, $s^*$.  The effects included in the final best subsets model of size $s^*$ are declared active. 
\item Regression with LB is described in Section~\ref{sec:CV}. Recall that the first step requires an estimate for $\sigma^2$ using the full model.  However, for $n=20$ and $m=7$, we cannot fit a full regression model containing all ME and 2FI's.  Rather, we utilize a ridge regression-based estimate of the error variance as proposed by \cite{liu2020estimation}. This estimate is obtained via the \emph{RidgeVar} package available for download here: https://github.com/xliusufe/RidgeVar. No other modifications were made to the LB procedure outlined in Section~\ref{fig:CV}.
\end{itemize}

\noindent{For each of the simulated response vectors, design types, and analysis methods, we measured the proportion of truly active effects identified as active (power) and the proportion of truly inactive effects identified as active (type 1 error).} 

\begin{figure}[H]
    \centering
    \includegraphics[width=0.9\linewidth]{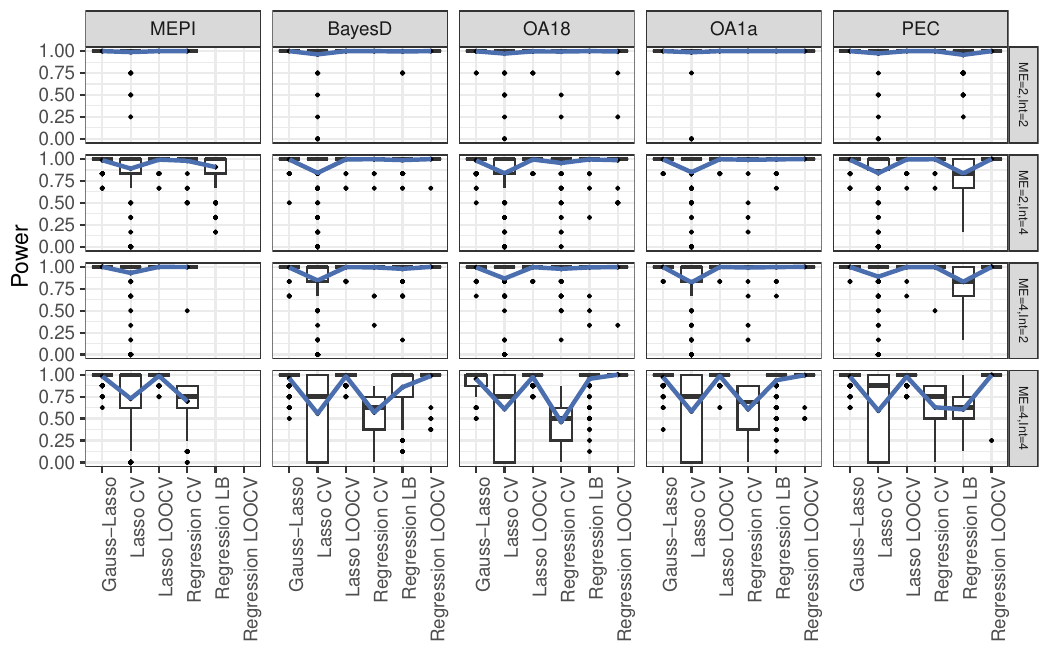}
    \caption{Power across all $n=20$ and $m=7$ design types across the combinations of main effects (ME) and two-factor interactions (Int) in the true model.The blue line indicates the average value. }
    \label{fig:n20_power}
\end{figure}

Figure~\ref{fig:n20_power} displays the power for each analysis method. We see that for all designs and analysis methods the power is high for the scenario where the true model contains just two active main effects and two active interactions.  As the true model increases in complexity (moving down the rows in Figure~\ref{fig:n20_power}) we see a decrease in power, particularly for Lasso CV and Regression CV.  Note: there are some missing values for the MEPI design, indicating that regression with LOOCV and LB failed for that design. It is unclear as to why this occurred, but may be related to issues with the best subsets algorithm in the R $leaps$ package.  Overall, the methods using LOOCV, along with the Gauss-Lasso, seem to have the highest power. We also observed a power drop for the PEC design using regression LB.

\begin{figure}[H]
    \centering
    \includegraphics[width=0.9\linewidth]{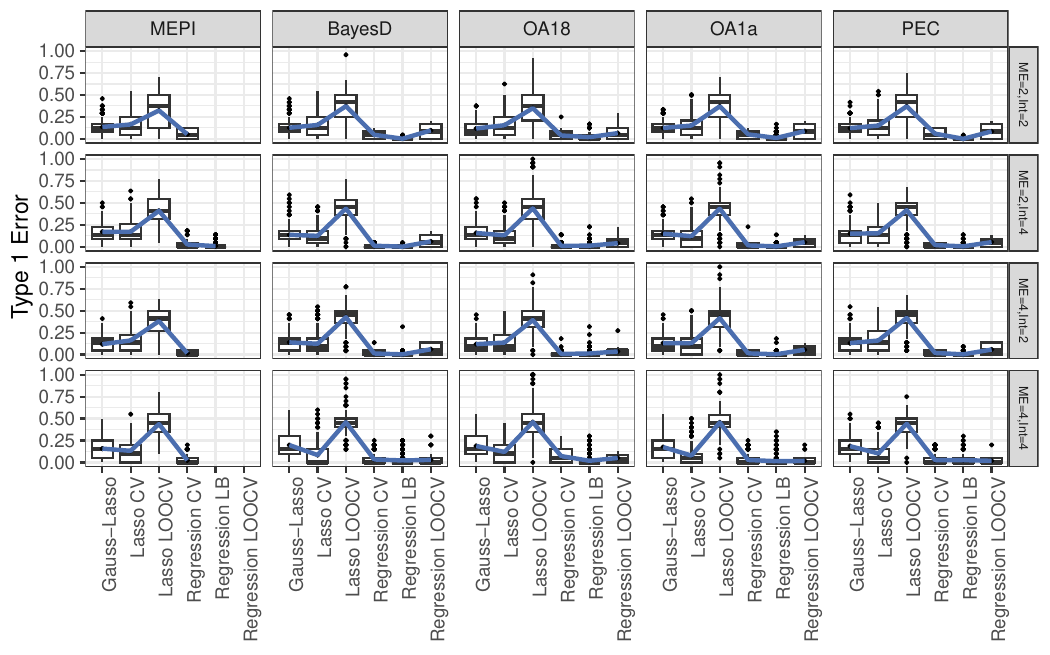}
    \caption{Type 1 error rate across all $n=20$ and $m=7$ design types across the combinations of main effects (ME) and two-factor interactions (Int) in the true model. The blue line indicates the average value.}
    \label{fig:n20_type1}
\end{figure}

Power alone is inadequate to judge these methods, so Figure~\ref{fig:n20_type1} displays their type 1 error rates.  Unlike power, the type 1 error rates are fairly consistent as the true model increases in complexity. Two of the highest-powered methods, lasso using LOOCV and the Gauss-Lasso, also have the highest type 1 error rates. This leaves regression with LOOCV as the overall best performer when accounting for both power and type 1 error rates. Of the lasso-based methods, the Gauss-Lasso provides the best balance of power and type 1 error.

\subsection{Supersaturated Screening} \label{sec:ssd_screen}

In this Section we explore a more extreme case of screening using supersaturated designs (SSD), where the number of available runs is less than the number of factors; that is, $n<m$. We revisit the simulation structure and designs used in \cite{marley2010comparison}.  That is, three different ($n$, $m$) SSD sizes: (14, 24), (12, 26) and (18, 22), and two different construction criteria: $E(s^2)$-optimality \citep{booth1962some} and Bayesian-D optimality \citep{jones2008bayesian}.  In addition to those designs we include SSDs constructed using the $Var(s+)$ criterion \citep{weese2017criterion}.  Following the protocol of \cite{marley2010comparison} we generate a true response according to $y=X\betavec+\epsilon$ where $\epsilon\sim N(0,I)$ and $X$ is the model matrix containing only the intercept and active MEs. We generate models using four combinations of effect sparsity and magnitude of the true effects where $a$ is the number of truly active effects and $\mu$ is the magnitude of effects:

\begin{itemize}
    \item Scenario 1: $a=3$, $\mu$=3
    \item Scenario 2: $a=4$, $\mu=4$
    \item Scenario 3: $a=6$, $\mu=6$
    \item Scenario 4: $a=9$, $\mu=\{10, 8, 5, 3, 2, 2, 2, 2, 2\}$
\end{itemize}

\noindent For each of the $a$ and $\mu$ scenarios, we generated 250 response vectors using the three different SSDs sizes for each of the three different SSD construction methods. For each response vector we calculated the power and the type 1 error rate using Lasso CV and LOOCV, Regression CV and LOOCV, the Gauss-Lasso, and Regression LB.  All methods are implemented as described previously.

\begin{figure}[H]
    \centering
    \includegraphics[width=0.8\linewidth]{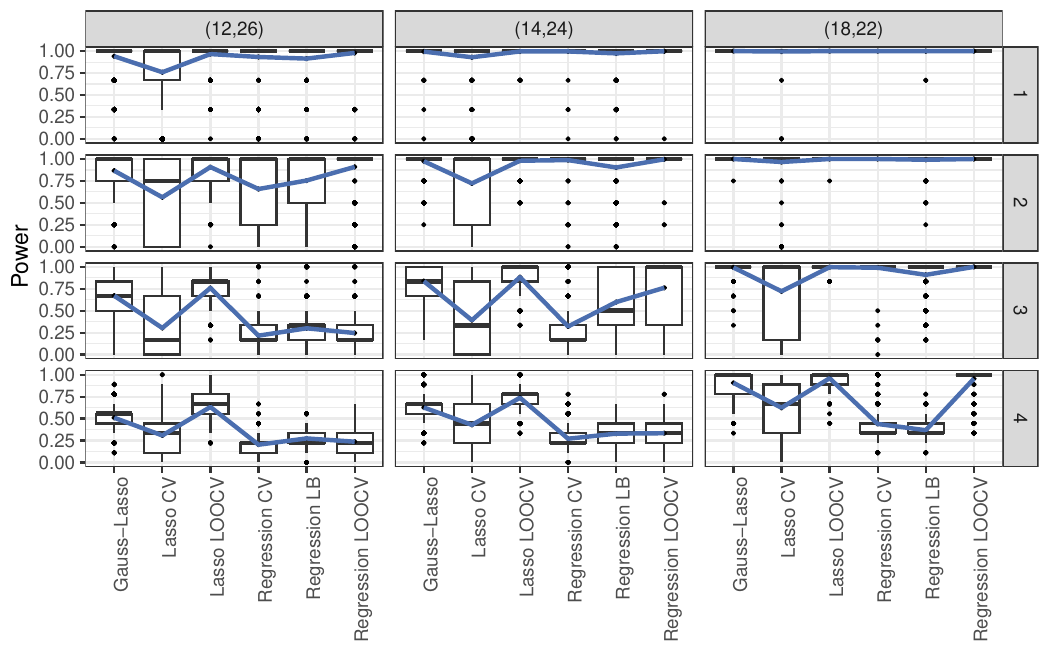}
    \caption{Power for each of the simulation scenarios (1-4) for each analysis method and each SSD size. The blue line indicates the average value.}
    \label{fig:ssd_power}
\end{figure}

\begin{figure}[H]
    \centering
    \includegraphics[width=0.8\linewidth]{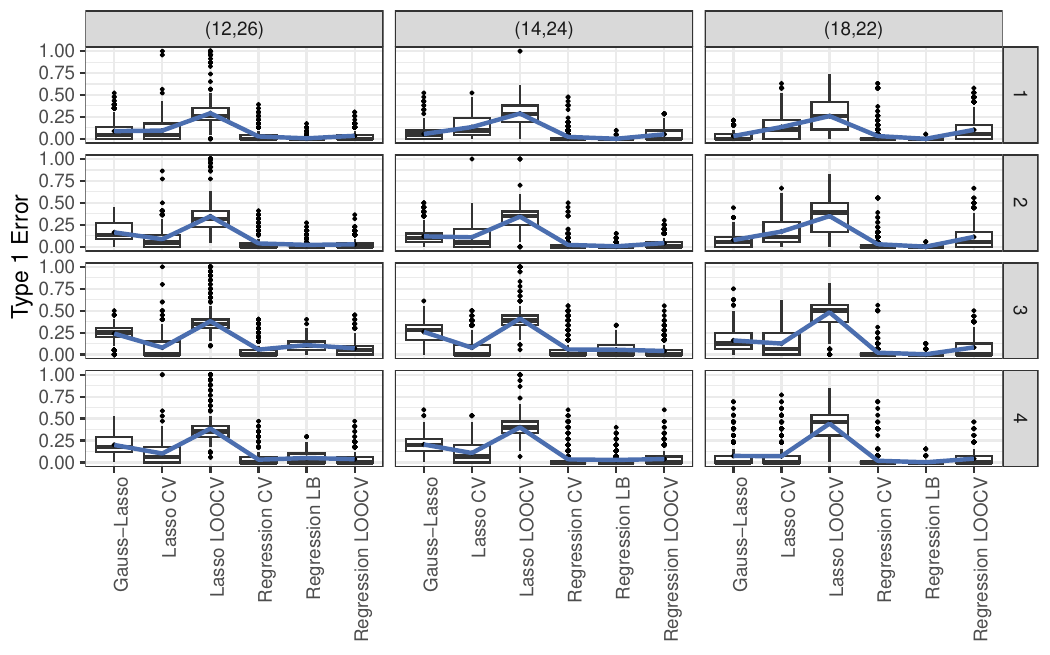}
    \caption{Type 1 error rate for each of the simulation scenarios (1-4) for each analysis method and each SSD size. The blue line indicates the average value.}
    \label{fig:ssd_type1}
\end{figure}

Figures~\ref{fig:ssd_power} and \ref{fig:ssd_type1} display the power and type 1 error across the three SSD sizes and for all four simulation scenarios.  Scenario 1 has higher power for all methods, and we see the power decreasing as the sparsity and effect size of the true model decrease (as expected).  We also notice that the power for the most supersaturated case $(12,26)$ is the lowest.  Lasso CV has the worst power across all scenarios and sizes; lasso LOOCV exhibits the highest power, but also the highest type 1 error rates. Of the lasso methods, Gauss-Lasso achieves the a desirable balance between power and type 1 error. Regression CV and regression LB show similar performance with good type I error control, while regression LOOCV generally yields improved results.  The results across the different design types are very similar, see Figures 1 and 2 in the supplementary materials.  
Overall, the Gauss-Lasso and Regression LOOCV are probably the best choices for SSD screening.

\section{Discussion} \label{sec:discussion}

The goal of this study was to assess the use of CV in the analysis of designed experiments, given the recent growth in DOE+ML and the literature's cautions about using it in the context of DOE. 
Although prior literature cautions against combining CV with designed experiments—suggesting it may undermine the structure of small designs—our findings do not support this concern. In practice, CV-based methods were often competitive, with regression using LOOCV performing particularly well across both small and large designs. This holds true for prediction in response surface models and for screening experiments. Motivated by Breiman’s critique of CV in design settings, we also evaluated his proposed alternative, the little bootstrap. While it performed adequately in the response surface context, it was consistently outperformed by other methods in screening scenarios.


More specific conclusions follow. If prediction is the ultimate analysis goal, the full regression model is a robust choice regardless of the nature of the true model. Of the model selection methods, regression with 5-fold CV or LOOCV perform well. A random forest model with LOOCV performs competitively in this context, but using 5-fold CV may lead to severe overfitting if the true model is relatively simple. For prediction, we find the implementation of regression with LB often has more variable performance than regression with CV or LOOCV. This could be due to having too few $n_{bootstrap}$ iterations; further investigation is warranted.

Although CV and similar methods are traditionally used for prediction tasks, we find that regression with LOOCV is capable of effective model selection in the case of screening as well. We found that regression with LOOCV, implemented using \emph{regsubsets}, may encounter implementation issues for certain designs, as we did with the MEPI design (Figures \ref{fig:n20_power} and \ref{fig:n20_type1}). Because we did not have any problems fitting the SSDs, we suspect that the cause of the problem is due to complex aliasing among two-factor interaction columns induced by the $(20,7)$ MEPI design.


We also note the relatively poor performance of Lasso with 5-fold CV and Lasso with LOOCV. The implementation of Gauss-Lasso uses thresholding to eliminate columns with very small lasso estimates for any given value of $\lambda$.  It is possible that using thresholding with Lasso CV/LOOCV could improve the type 1 error rates we encountered. It is also possible that if the Lasso CV/LOOCV model selection procedure was augmented with a subjective, visual evaluation of the profile plot, its performance would be improved.

Finally, the empirical study of the little bootstrap suggests several future research directions. First, could the implementation of the LB be tuned in order to improve its performance? For instance, it requires the estimation of $\sigma^{2}$, which we achieved in the screening experiment setting by using the ridge regression-based method of \citet{liu2020estimation}. Might some alternatives provide better performance? There are also several tuning parameters in the LB, including the variance scaling parameter $t$ and the number of bootstrap samples $n_{bootstrap}$. More generally, the use of the ridge regression-based estimator of $\sigma^{2}$ in designed experiments---especially those like SSDs that do not have such an estimator readily available---has not been explored and could be important future work.

\section*{Data Availability}
The data that support the findings of this study are available from the corresponding author, Maria Weese, upon reasonable request.

\section*{Supplementary Files}

Designs, code and supplementary graphics can be found at github.com/weeseml/CVandDOE.

\bibliographystyle{asa} 
\bibliography{references.bib}

\begin{thebibliography}{77}
\newcommand{\enquote}[1]{``#1''}
\expandafter\ifx\csname natexlab\endcsname\relax\def\natexlab#1{#1}\fi

\bibitem[{A.~Ramezan et~al.(2019)A.~Ramezan, A.~Warner, and E.~Maxwell}]{a2019evaluation}
A.~Ramezan, C., A.~Warner, T., and E.~Maxwell, A. (2019), \enquote{Evaluation of sampling and cross-validation tuning strategies for regional-scale machine learning classification,} \textit{Remote Sensing}, 11, 185.

\bibitem[{Abedpour et~al.(2023)Abedpour, Moghaddas, Sori, and Alizadeh}]{abedpour2023experimental}
Abedpour, H., Moghaddas, J., Sori, A., and Alizadeh, R. (2023), \enquote{Experimental study and machine learning simulation of Pb (II) separation from aqueous solutions via a nanocomposite adsorbent,} \textit{Journal of the Taiwan Institute of Chemical Engineers}, 147, 104923.

\bibitem[{Akaike(1974)}]{akaikeAIC}
Akaike, H. (1974), \enquote{A new look at the statistical model identification,} \textit{IEEE transactions on automatic control}, 19, 716--723.

\bibitem[{Arboretti et~al.(2022{\natexlab{a}})Arboretti, Ceccato, Pegoraro, and Salmaso}]{arboretti2022designQREI}
Arboretti, R., Ceccato, R., Pegoraro, L., and Salmaso, L. (2022{\natexlab{a}}), \enquote{Design choice and machine learning model performances,} \textit{Quality and Reliability Engineering International}, 38, 3357--3378.

\bibitem[{Arboretti et~al.(2022{\natexlab{b}})Arboretti, Ceccato, Pegoraro, and Salmaso}]{arboretti2022design}
--- (2022{\natexlab{b}}), \enquote{Design of Experiments and machine learning for product innovation: A systematic literature review,} \textit{Quality and Reliability Engineering International}, 38, 1131--1156.

\bibitem[{Arboretti et~al.(2021)Arboretti, Ceccato, Pegoraro, Salmaso, Housmekerides, Spadoni, Pierangelo, Quaggia, Tveit, and Vianello}]{arboretti2021machine}
Arboretti, R., Ceccato, R., Pegoraro, L., Salmaso, L., Housmekerides, C., Spadoni, L., Pierangelo, E., Quaggia, S., Tveit, C., and Vianello, S. (2021), \enquote{Machine learning and design of experiments with an application to product innovation in the chemical industry,} \textit{Journal of Applied Statistics}, 1--26.

\bibitem[{Arboretti et~al.(2022{\natexlab{c}})Arboretti, Ceccato, Pegoraro, Salmaso, Housmekerides, Spadoni, Pierangelo, Quaggia, Tveit, and Vianello}]{arboretti2022machine}
--- (2022{\natexlab{c}}), \enquote{Machine learning and design of experiments with an application to product innovation in the chemical industry,} \textit{Journal of Applied Statistics}, 49, 2674--2699.

\bibitem[{Arlot and Celisse(2010)}]{arlot2010survey}
Arlot, S. and Celisse, A. (2010), \enquote{A survey of cross-validation procedures for model selection,} \textit{Statistics Surveys}, 4, 40--79.

\bibitem[{Bates et~al.(2024)Bates, Hastie, and Tibshirani}]{bates2024cross}
Bates, S., Hastie, T., and Tibshirani, R. (2024), \enquote{Cross-validation: what does it estimate and how well does it do it?} \textit{Journal of the American Statistical Association}, 119, 1434--1445.

\bibitem[{Blum et~al.(1999)Blum, Kalai, and Langford}]{blum1999beating}
Blum, A., Kalai, A., and Langford, J. (1999), \enquote{Beating the hold-out: Bounds for k-fold and progressive cross-validation,} in \textit{Proceedings of the twelfth annual conference on Computational learning theory}, pp. 203--208.

\bibitem[{Booth and Cox(1962)}]{booth1962some}
Booth, K.~H. and Cox, D.~R. (1962), \enquote{Some systematic supersaturated designs,} \textit{Technometrics}, 4, 489--495.

\bibitem[{Box and Draper(2007)}]{box2007response}
Box, G.~E. and Draper, N.~R. (2007), \textit{Response Surfaces, Mixtures, and Ridge Analyses}, Springer.

\bibitem[{Box and Wilson(1951)}]{Box1951}
Box, G. E.~P. and Wilson, K.~B. (1951), \enquote{On the Experimental Attainment of Optimum Conditions,} \textit{Journal of the Royal Statistical Society. Series B, Methodological}, 13, 1--45.

\bibitem[{Breiman(1992)}]{breiman1992little}
Breiman, L. (1992), \enquote{The little bootstrap and other methods for dimensionality selection in regression: X-fixed prediction error,} \textit{Journal of the American Statistical Association}, 87, 738--754.

\bibitem[{Breiman(1996)}]{breiman1996heuristics}
--- (1996), \enquote{Heuristics of instability and stabilization in model selection,} \textit{The annals of statistics}, 24, 2350--2383.

\bibitem[{Dietterich(1998)}]{dietterich1998approximate}
Dietterich, T.~G. (1998), \enquote{Approximate statistical tests for comparing supervised classification learning algorithms,} \textit{Neural computation}, 10, 1895--1923.

\bibitem[{Dragulji{\'c} et~al.(2014)Dragulji{\'c}, Woods, Dean, Lewis, and Vine}]{draguljic2014screening}
Dragulji{\'c}, D., Woods, D.~C., Dean, A.~M., Lewis, S.~M., and Vine, A.-J.~E. (2014), \enquote{Screening strategies in the presence of interactions,} \textit{Technometrics}, 56, 1--1.

\bibitem[{Dropka et~al.(2022)Dropka, Tang, Chappa, and Holena}]{dropka2022smart}
Dropka, N., Tang, X., Chappa, G.~K., and Holena, M. (2022), \enquote{Smart Design of Cz-Ge Crystal Growth Furnace and Process,} \textit{Crystals}, 12, 1764.

\bibitem[{Duarte and Wainer(2017)}]{duarte2017empirical}
Duarte, E. and Wainer, J. (2017), \enquote{Empirical comparison of cross-validation and internal metrics for tuning SVM hyperparameters,} \textit{Pattern Recognition Letters}, 88, 6--11.

\bibitem[{DuMouchel and Jones(1994)}]{dumouchel1994simple}
DuMouchel, W. and Jones, B. (1994), \enquote{A simple Bayesian modification of D-optimal designs to reduce dependence on an assumed model,} \textit{Technometrics}, 36, 37--47.

\bibitem[{Elkatatny et~al.(2023)Elkatatny, Alsharekh, Alateyah, El-Sanabary, Nassef, Kamel, Alawad, BaQais, El-Garaihy, and Kouta}]{elkatatny2023optimizing}
Elkatatny, S., Alsharekh, M.~F., Alateyah, A.~I., El-Sanabary, S., Nassef, A., Kamel, M., Alawad, M.~O., BaQais, A., El-Garaihy, W.~H., and Kouta, H. (2023), \enquote{Optimizing the Powder Metallurgy Parameters to Enhance the Mechanical Properties of Al-4Cu/xAl2O3 Composites Using Machine Learning and Response Surface Approaches,} \textit{Applied Sciences}, 13, 7483.

\bibitem[{Escribano-Garc{\'\i}a et~al.(2014)Escribano-Garc{\'\i}a, Lostado-Lorza, Fern{\'a}ndez-Mart{\'\i}nez, Villanueva-Rold{\'a}n, and Mac~Donald}]{escribano2014improvement}
Escribano-Garc{\'\i}a, R., Lostado-Lorza, R., Fern{\'a}ndez-Mart{\'\i}nez, R., Villanueva-Rold{\'a}n, P., and Mac~Donald, B.~J. (2014), \enquote{Improvement in manufacturing welded products through multiple response surface methodology and data mining techniques,} in \textit{International Joint Conference SOCO’14-CISIS’14-ICEUTE’14: Bilbao, Spain, June 25th-27th, 2014, Proceedings}, Springer, pp. 301--310.

\bibitem[{Friedman et~al.(2021)Friedman, Hastie, Tibshirani, Narasimhan, Tay, Simon, and Qian}]{friedman2021package}
Friedman, J., Hastie, T., Tibshirani, R., Narasimhan, B., Tay, K., Simon, N., and Qian, J. (2021), \enquote{Package ‘glmnet’,} \textit{CRAN R Repositary}, 595.

\bibitem[{Geisser(1975)}]{geisser1975predictive}
Geisser, S. (1975), \enquote{The predictive sample reuse method with applications,} \textit{Journal of the American statistical Association}, 70, 320--328.

\bibitem[{Ghalandari et~al.(2019)Ghalandari, Ziamolki, Mosavi, Shamshirband, Chau, and Bornassi}]{ghalandari2019aeromechanical}
Ghalandari, M., Ziamolki, A., Mosavi, A., Shamshirband, S., Chau, K.-W., and Bornassi, S. (2019), \enquote{Aeromechanical optimization of first row compressor test stand blades using a hybrid machine learning model of genetic algorithm, artificial neural networks and design of experiments,} \textit{Engineering Applications of Computational Fluid Mechanics}, 13, 892--904.

\bibitem[{Ginige et~al.(2021)Ginige, Song, Olsen, Luber, Yavuz, and Buriak}]{ginige2021solvent}
Ginige, G., Song, Y., Olsen, B.~C., Luber, E.~J., Yavuz, C.~T., and Buriak, J.~M. (2021), \enquote{Solvent vapor annealing, defect analysis, and optimization of self-assembly of block copolymers using machine learning approaches,} \textit{ACS Applied Materials \& Interfaces}, 13, 28639--28649.

\bibitem[{Gorriz et~al.(2024)Gorriz, Clemente, Segovia, Ramirez, Ortiz, and Suckling}]{gorriz2024kfoldcrossvalidationbest}
Gorriz, J.~M., Clemente, R.~M., Segovia, F., Ramirez, J., Ortiz, A., and Suckling, J. (2024), \enquote{Is K-fold cross validation the best model selection method for Machine Learning?} .

\bibitem[{James et~al.(2013)James, Witten, Hastie, Tibshirani, et~al.}]{james2013introduction}
James, G., Witten, D., Hastie, T., Tibshirani, R., et~al. (2013), \textit{An introduction to statistical learning}, vol. 112, Springer.

\bibitem[{Jones et~al.(2008)Jones, Lin, and Nachtsheim}]{jones2008bayesian}
Jones, B., Lin, D.~K., and Nachtsheim, C.~J. (2008), \enquote{Bayesian D-optimal supersaturated designs,} \textit{Journal of Statistical Planning and Inference}, 138, 86--92.

\bibitem[{Joseph et~al.(2015)Joseph, Gul, and Ba}]{joseph2015maximum}
Joseph, V.~R., Gul, E., and Ba, S. (2015), \enquote{Maximum projection designs for computer experiments,} \textit{Biometrika}, 102, 371--380.

\bibitem[{Joseph et~al.(2020)Joseph, Gul, and Ba}]{joseph2020designing}
--- (2020), \enquote{Designing computer experiments with multiple types of factors: The MaxPro approach,} \textit{Journal of Quality Technology}, 52, 343--354.

\bibitem[{Khan et~al.(2022)Khan, Wahab, Hussain, Khan, and Rashid}]{khan2022multi}
Khan, H., Wahab, F., Hussain, S., Khan, S., and Rashid, M. (2022), \enquote{Multi-object optimization of Navy-blue anodic oxidation via response surface models assisted with statistical and machine learning techniques,} \textit{Chemosphere}, 291, 132818.

\bibitem[{Kohavi et~al.(1995)}]{kohavi1995study}
Kohavi, R. et~al. (1995), \enquote{A study of cross-validation and bootstrap for accuracy estimation and model selection,} in \textit{Ijcai}, Montreal, Canada, vol.~14, pp. 1137--1145.

\bibitem[{Krstajic et~al.(2014)Krstajic, Buturovic, Leahy, and Thomas}]{krstajic2014cross}
Krstajic, D., Buturovic, L.~J., Leahy, D.~E., and Thomas, S. (2014), \enquote{Cross-validation pitfalls when selecting and assessing regression and classification models,} \textit{Journal of cheminformatics}, 6, 1--15.

\bibitem[{Kuhn et~al.(2013)Kuhn, Johnson, et~al.}]{kuhn2013applied}
Kuhn, M., Johnson, K., et~al. (2013), \textit{Applied predictive modeling}, vol.~26, Springer.

\bibitem[{Lemkus et~al.(2021)Lemkus, Gotwalt, Ramsey, and Weese}]{lemkus2021self}
Lemkus, T., Gotwalt, C., Ramsey, P., and Weese, M.~L. (2021), \enquote{Self-validated ensemble models for design of experiments,} \textit{Chemometrics and Intelligent Laboratory Systems}, 219, 104439.

\bibitem[{Li and Nachtsheim(2000)}]{li2000model}
Li, W. and Nachtsheim, C.~J. (2000), \enquote{Model-robust factorial designs,} \textit{Technometrics}, 42, 345--352.

\bibitem[{Lin(2021)}]{lin2021forward}
Lin, C.-Y. (2021), \enquote{Forward stepwise random forest analysis for experimental designs,} \textit{Journal of Quality Technology}, 53, 488--504.

\bibitem[{Liu et~al.(2020)Liu, Zheng, and Feng}]{liu2020estimation}
Liu, X., Zheng, S., and Feng, X. (2020), \enquote{Estimation of error variance via ridge regression,} \textit{Biometrika}, 107, 481--488.

\bibitem[{Loeppky et~al.(2007)Loeppky, Sitter, and Tang}]{loeppky2007nonregular}
Loeppky, J.~L., Sitter, R.~R., and Tang, B. (2007), \enquote{Nonregular designs with desirable projection properties,} \textit{Technometrics}, 49, 454--467.

\bibitem[{Lumley and Lumley(2013)}]{lumley2013package}
Lumley, T. and Lumley, M.~T. (2013), \enquote{Package ‘leaps’,} \textit{Regression subset selection. Thomas Lumley Based on Fortran Code by Alan Miller. Available online: http://CRAN. R-project. org/package= leaps (Accessed on 18 March 2018)}.

\bibitem[{Marley and Woods(2010)}]{marley2010comparison}
Marley, C.~J. and Woods, D.~C. (2010), \enquote{A comparison of design and model selection methods for supersaturated experiments,} \textit{Computational Statistics \& Data Analysis}, 54, 3158--3167.

\bibitem[{Mathew et~al.(2020)Mathew, Karandikar, and Kulkarni}]{mathew2020modeling}
Mathew, S., Karandikar, P.~B., and Kulkarni, N.~R. (2020), \enquote{Modeling and Optimization of a Jackfruit Seed-Based Supercapacitor Electrode Using Machine Learning,} \textit{Chemical Engineering \& Technology}, 43, 1765--1773.

\bibitem[{McDaniel and Ankenman(2000)}]{mcdaniel2000response}
McDaniel, W.~R. and Ankenman, B.~E. (2000), \enquote{A response surface test bed,} \textit{Quality and Reliability Engineering International}, 16, 363--372.

\bibitem[{Mee et~al.(2017)Mee, Schoen, and Edwards}]{mee2017selecting}
Mee, R.~W., Schoen, E.~D., and Edwards, D.~J. (2017), \enquote{Selecting an orthogonal or nonorthogonal two-level design for screening,} \textit{Technometrics}, 59, 305--318.

\bibitem[{Myers et~al.(2016)Myers, Montgomery, and Anderson-Cook}]{myers2016response}
Myers, R.~H., Montgomery, D.~C., and Anderson-Cook, C.~M. (2016), \textit{Response surface methodology: process and product optimization using designed experiments}, John Wiley \& Sons.

\bibitem[{Nikita et~al.(2023)Nikita, Sharma, Fahmi, and Rathore}]{nikita2023process}
Nikita, S., Sharma, R., Fahmi, J., and Rathore, A.~S. (2023), \enquote{Process optimization using machine learning enhanced design of experiments (DOE): ranibizumab refolding as a case study,} \textit{Reaction Chemistry \& Engineering}, 8, 592--603.

\bibitem[{Ockuly et~al.(2017)Ockuly, Weese, Smucker, Edwards, and Chang}]{ockuly2017response}
Ockuly, R.~A., Weese, M.~L., Smucker, B.~J., Edwards, D.~J., and Chang, L. (2017), \enquote{Response surface experiments: A meta-analysis,} \textit{Chemometrics and Intelligent Laboratory Systems}, 164, 64--75.

\bibitem[{Pinto et~al.(2019)Pinto, de~Azevedo, Oliveira, and von Stosch}]{pinto2019bootstrap}
Pinto, J., de~Azevedo, C.~R., Oliveira, R., and von Stosch, M. (2019), \enquote{A bootstrap-aggregated hybrid semi-parametric modeling framework for bioprocess development,} \textit{Bioprocess and biosystems engineering}, 42, 1853--1865.

\bibitem[{Probst et~al.(2019)Probst, Boulesteix, and Bischl}]{probst2019tunability}
Probst, P., Boulesteix, A.-L., and Bischl, B. (2019), \enquote{Tunability: Importance of hyperparameters of machine learning algorithms,} \textit{Journal of Machine Learning Research}, 20, 1--32.

\bibitem[{Rabiee et~al.(2023)Rabiee, Tahmasbi, and Qasemi}]{rabiee2023experimental}
Rabiee, A.~H., Tahmasbi, V., and Qasemi, M. (2023), \enquote{Experimental evaluation, modeling and sensitivity analysis of temperature and cutting force in bone micro-milling using support vector regression and EFAST methods,} \textit{Engineering Applications of Artificial Intelligence}, 120, 105874.

\bibitem[{Raghavan et~al.(2023)Raghavan, Jain, Chowdhury, Sakuma, Doll, Biesheuvel, Van~Borkulo, and Boughorbel}]{raghavan2023methodology}
Raghavan, S., Jain, A., Chowdhury, P.~R., Sakuma, K., Doll, R., Biesheuvel, K., Van~Borkulo, J., and Boughorbel, F. (2023), \enquote{A Methodology to Optimize Laser Dicing Parameters to Maximize Dicing Quality Through Machine Learning,} in \textit{2023 IEEE 73rd Electronic Components and Technology Conference (ECTC)}, IEEE, pp. 139--146.

\bibitem[{Raschka(2020)}]{raschka2020modelevaluationmodelselection}
Raschka, S. (2020), \enquote{Model Evaluation, Model Selection, and Algorithm Selection in Machine Learning,} .

\bibitem[{Ratnavel et~al.(2022)Ratnavel, Viswanath, Subramanian, Selvaraj, Prahasam, and Siddharth}]{ratnavel2022predicting}
Ratnavel, R., Viswanath, S., Subramanian, J., Selvaraj, V.~K., Prahasam, V., and Siddharth, S. (2022), \enquote{Predicting the Optimal Input Parameters for the Desired Print Quality Using Machine Learning,} \textit{Micromachines}, 13, 2231.

\bibitem[{Raz et~al.(2018)Raz, Wood, Mockus, DeLaurentis, and Llinas}]{raz2018identifying}
Raz, A.~K., Wood, P., Mockus, L., DeLaurentis, D.~A., and Llinas, J. (2018), \enquote{Identifying interactions for information fusion system design using machine learning techniques,} in \textit{2018 21st International Conference on Information Fusion (FUSION)}, IEEE, pp. 226--233.

\bibitem[{Rebollo et~al.(2022)Rebollo, Oyoun, Corvis, El-Hammadi, Saubamea, Andrieux, Mignet, and Alhareth}]{rebollo2022microfluidic}
Rebollo, R., Oyoun, F., Corvis, Y., El-Hammadi, M.~M., Saubamea, B., Andrieux, K., Mignet, N., and Alhareth, K. (2022), \enquote{Microfluidic manufacturing of liposomes: development and optimization by design of experiment and machine learning,} \textit{ACS Applied Materials \& Interfaces}, 14, 39736--39745.

\bibitem[{Ronchetti et~al.(1997)Ronchetti, Field, and Blanchard}]{ronchetti1997robust}
Ronchetti, E., Field, C., and Blanchard, W. (1997), \enquote{Robust linear model selection by cross-validation,} \textit{Journal of the American Statistical Association}, 92, 1017--1023.

\bibitem[{Saxena et~al.(2021)Saxena, Roman, Robu, Flynn, and Pecht}]{saxena2021battery}
Saxena, S., Roman, D., Robu, V., Flynn, D., and Pecht, M. (2021), \enquote{Battery stress factor ranking for accelerated degradation test planning using machine learning,} \textit{energies}, 14, 723.

\bibitem[{Schratz et~al.(2019)Schratz, Muenchow, Iturritxa, Richter, and Brenning}]{schratz2019hyperparameter}
Schratz, P., Muenchow, J., Iturritxa, E., Richter, J., and Brenning, A. (2019), \enquote{Hyperparameter tuning and performance assessment of statistical and machine-learning algorithms using spatial data,} \textit{Ecological Modelling}, 406, 109--120.

\bibitem[{Schwarz(1978)}]{schwarz1978estimating}
Schwarz, G. (1978), \enquote{Estimating the dimension of a model,} \textit{The annals of statistics}, 461--464.

\bibitem[{Shao(1993)}]{shao1993linear}
Shao, J. (1993), \enquote{Linear model selection by cross-validation,} \textit{Journal of the American statistical Association}, 88, 486--494.

\bibitem[{Smucker et~al.(2021)Smucker, Edwards, and Weese}]{smucker2021response}
Smucker, B.~J., Edwards, D.~J., and Weese, M.~L. (2021), \enquote{Response surface models: To reduce or not to reduce?} \textit{Journal of Quality Technology}, 53, 197--216.

\bibitem[{Stone(1974)}]{stone1974cross}
Stone, M. (1974), \enquote{Cross-validatory choice and assessment of statistical predictions,} \textit{Journal of the royal statistical society: Series B (Methodological)}, 36, 111--133.

\bibitem[{Varma and Simon(2006)}]{varma2006bias}
Varma, S. and Simon, R. (2006), \enquote{Bias in error estimation when using cross-validation for model selection,} \textit{BMC bioinformatics}, 7, 1--8.

\bibitem[{Varoquaux et~al.(2017)Varoquaux, Raamana, Engemann, Hoyos-Idrobo, Schwartz, and Thirion}]{varoquaux2017assessing}
Varoquaux, G., Raamana, P.~R., Engemann, D.~A., Hoyos-Idrobo, A., Schwartz, Y., and Thirion, B. (2017), \enquote{Assessing and tuning brain decoders: cross-validation, caveats, and guidelines,} \textit{NeuroImage}, 145, 166--179.

\bibitem[{Weese et~al.(2017)Weese, Edwards, and Smucker}]{weese2017criterion}
Weese, M.~L., Edwards, D.~J., and Smucker, B.~J. (2017), \enquote{A criterion for constructing powerful supersaturated designs when effect directions are known,} \textit{Journal of Quality Technology}, 49, 265--277.

\bibitem[{Weese et~al.(2021)Weese, Stallrich, Smucker, and Edwards}]{weese2021strategies}
Weese, M.~L., Stallrich, J.~W., Smucker, B.~J., and Edwards, D.~J. (2021), \enquote{Strategies for supersaturated screening: Group orthogonal and constrained var (s) designs,} \textit{Technometrics}, 63, 443--455.

\bibitem[{Wiemer et~al.(2019)Wiemer, Drowatzky, and Ihlenfeldt}]{wiemer2019data}
Wiemer, H., Drowatzky, L., and Ihlenfeldt, S. (2019), \enquote{Data mining methodology for engineering applications (DMME)—A holistic extension to the CRISP-DM model,} \textit{Applied Sciences}, 9, 2407.

\bibitem[{Wright and Ziegler(2015)}]{wright2015ranger}
Wright, M.~N. and Ziegler, A. (2015), \enquote{ranger: A fast implementation of random forests for high dimensional data in C++ and R,} \textit{arXiv preprint arXiv:1508.04409}.

\bibitem[{Xu and Liang(2001)}]{xu2001monte}
Xu, Q.-S. and Liang, Y.-Z. (2001), \enquote{Monte Carlo cross validation,} \textit{Chemometrics and Intelligent Laboratory Systems}, 56, 1--11.

\bibitem[{Yang(2007)}]{yang2007consistency}
Yang, Y. (2007), \enquote{Consistency of cross validation for comparing regression procedures,} \textit{Annals of Statistics}, 35, 2450--2473.

\bibitem[{Yates et~al.(2023)Yates, Aandahl, Richards, and Brook}]{yates2023cross}
Yates, L.~A., Aandahl, Z., Richards, S.~A., and Brook, B.~W. (2023), \enquote{Cross validation for model selection: a review with examples from ecology,} \textit{Ecological Monographs}, 93, e1557.

\bibitem[{Yuan et~al.(2007)Yuan, Joseph, and Lin}]{yuan2007efficient}
Yuan, M., Joseph, V.~R., and Lin, Y. (2007), \enquote{An efficient variable selection approach for analyzing designed experiments,} \textit{Technometrics}, 49, 430--439.

\bibitem[{Zalkikar et~al.(2022)Zalkikar, Nepal, Husin, Yadav, and Banerjee}]{zalkikar2022predictive}
Zalkikar, A., Nepal, B., Husin, H., Yadav, O., and Banerjee, A. (2022), \enquote{Predictive Analysis of Fluid-Hammer Effect on LNG Regasification System Pipeline Network,} in \textit{2022 Annual Reliability and Maintainability Symposium (RAMS)}, IEEE, pp. 1--6.

\bibitem[{Zhang(1993)}]{zhang1993model}
Zhang, P. (1993), \enquote{Model selection via multifold cross validation,} \textit{The annals of statistics}, 299--313.

\bibitem[{Zhang and Yang(2015)}]{zhang2015cross}
Zhang, Y. and Yang, Y. (2015), \enquote{Cross-validation for selecting a model selection procedure,} \textit{Journal of Econometrics}, 187, 95--112.

\bibitem[{Zhong et~al.(2010)Zhong, Fan, Yang, Verscheure, and Ren}]{zhong2010cross}
Zhong, E., Fan, W., Yang, Q., Verscheure, O., and Ren, J. (2010), \enquote{Cross validation framework to choose amongst models and datasets for transfer learning,} in \textit{Machine Learning and Knowledge Discovery in Databases: European Conference, ECML PKDD 2010, Barcelona, Spain, September 20-24, 2010, Proceedings, Part III 21}, Springer, pp. 547--562.

\end{thebibliography}

\end{document}